\documentclass[10pt, english,pra,aps,twocolumn,floatfix,superscriptaddress]{revtex4-2}
\usepackage{comment}
\usepackage{braket}
\usepackage{amsmath}
\usepackage{float}
\usepackage{tikz}
\usetikzlibrary{quantikz}
\usetikzlibrary{positioning,shadows,shapes,arrows}
\usepackage{dsfont}
\usepackage{geometry}
\usepackage{xcolor}
\usepackage{graphicx}
\usepackage{dcolumn}
\usepackage{bm}
\usepackage{hyperref}
\usepackage{microtype}
\usetikzlibrary{arrows}
\usetikzlibrary{arrows.meta}
\usetikzlibrary{shapes,arrows}
\usepackage{amsmath,bm,times}
\usepackage{tikz}
\usetikzlibrary{arrows.meta}

\begin{document}

\title{Quantum noise can enhance algorithmic cooling}

\author{Zahra Farahmand}
\affiliation{
 Department of Physics, Sharif University of Technology, Tehran, Iran
}
\author{Reyhaneh Aghaei Saem}
\affiliation{
 Department of Physics, Sharif University of Technology, Tehran, Iran
}
\author{Sadegh Raeisi}
\email{sraeisi@sharif.edu}
\affiliation{
 Department of Physics, Sharif University of Technology, Tehran, Iran
}

\begin{abstract}
Heat-Bath Algorithmic Cooling techniques (HBAC) are techniques that are used to purify a target element in a quantum system. These methods compress and transfer entropy away from the target element into auxiliary elements of the system. 
The performance of Algorithmic Cooling has been investigated under ideal noiseless conditions. However, realistic implementations are imperfect and for practical purposes, noise should be taken into account. Here we analyze Heat-Bath Algorithmic Cooling techniques under realistic noise models. 
Surprisingly, we find that noise can in some cases enhance the performance and improve the cooling limit of Heat-Bath Algorithmic Cooling techniques. We numerically simulate the noisy algorithmic cooling for the two optimal strategies, the Partner Pairing, and the Two-sort algorithms. We find that for both of them, in the presence of the generalized amplitude damping noise, the process converges and the asymptotic purity can be higher than the noiseless process. This opens up new avenues for increasing the purity beyond the heat-bath algorithmic cooling.
\end{abstract}


\maketitle
Most quantum applications require pure qubits. However, it is not always easy to prepare pure quantum states. For instance, for spin qubits, the states are often close to maximally mixed states. 

Heat-Bath Algorithmic Cooling techniques (HBAC) provide implementation-agnostic methods for the purification of qubits. These techniques exploit auxiliary qubits to increase the purity of target elements. 

Algorithmic cooling techniques work based on quantum compression algorithms. They compress the entropy of the system and transfer that away from the target elements and into the auxiliary elements in the system \cite{schulman1998scalable}. 

The initial algorithmic cooling techniques were limited by Shannon's bound for compression. 
In 2002, Boykin et al. proposed to use a heat-bath to reset the auxiliary qubits, and repeat the cooling process \cite{boykin2002algorithmic}. 
This turned the original cooling algorithms to iterative processes that could go beyond the Shannon's bound for compression. 
In each iteration, first, the compression operation would transfer the entropy away from the target qubit to the auxiliary ones, and then through the interaction with the heat-bath, the accumulated entropy would be transferred out of the system and to the heat-bath. 

In \cite{PhysRevLett.94.120501}, Shulman et al. introduced the Partner Pairing Algorithm (PPA) and showed that it is the optimal HBAC technique. PPA sorts the diagonal elements of the density matrix for compression. 
They showed that sort is the optimal compression operation for HBAC. 
They also proved that even this optimal technique cannot always converge to a completely pure state. 
This indicated that HBAC techniques are physically limited. 
Later, Raeisi and Mosca established the asymptotic state and found the cooling limit of HBAC techniques \cite{PhysRevLett.114.100404}. Recently the physical roots of the limit have been further investigated by Raeisi and it was shown that the limit is due to the unitarity of compression operations \cite{raeisi2021no}. Specifically, a no-go theorem was proved which shows that for two qubits, it is not possible to increase the purity beyond the purity of the individual qubits.   

One of the main challenges of PPA is that it is state-dependent and in each iteration, the complete information of the state of the system is required to determine the sort operation. 
This also means that the compression operation of PPA is constantly changing through the HBAC process. 
These make the implementation of PPA challenging and impractical. 
In \cite{raeisi2019novel}, Raeisi et al. introduced a new optimal HBAC technique. The new algorithm is called the ``Two-Sort Algorithmic Cooling'' (TSAC) technique and resolves these issues. They proved that this new method converges to the HBAC limit with a fixed compression operation. It means that TSAC can cool down the target qubits to the HBAC limit without knowing the state of the system. This was a significant step towards bringing HBAC to the realm of feasibility \cite{raeisi2020method}. 

One remaining caveat that has limited practical applications of HBAC is the assumption of ideal experimental conditions. Namely, most of the studies in this field investigate HBAC under noiseless conditions and do not take into account realistic imperfections. 
This problem is interesting from both fundamental and practical points of view. 
Fundamentally, it is interesting to know if a noisy implementation of HBAC would converge to a limit and also if the limit changes from the noiseless limit. 
Practically, it is unclear if HBAC can be helpful under realistic noisy implementation.

Here we investigate noisy HBAC and answer these questions. 
We find that for the generalized amplitude damping channel, which is one of the most relevant experimental noise models, noisy HBAC converges and interestingly, it also exceeds the HBAC limit. 
We also study the effect of other relevant noise models such as the depolarizing channel on the cooling algorithms. We find that only the generalized amplitude damping channel could enhance the HBAC limit. 

From a practical point of view, our results provide a more realistic assessment of HBAC. These findings show how HBAC is expected to perform under imperfect experimental conditions. 
In particular, for a specific experimental setting, the noise model can be characterized and our findings can help determine if and by how much HBAC is expected to help in increasing the purity.

We start by introducing our notation, the noise model, and our assumptions. 
We investigate the effect of quantum noise for both TSAC and PPA methods and identify the regimes where the noisy HBAC goes beyond the cooling limit of HBAC. 
We introduce some metrics to characterize the enhancement due to the noise. 
We find that the noise enhances PPA more effectively compared to TSAC, in the sense that, first, for identical noise parameters, the limit of noisy PPA is greater than the limit of noisy TSAC, and second, PPA is enhanced for a wider range of noise parameters compared to TSAC.

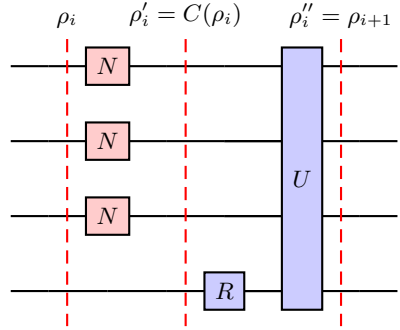
\begin{figure}[t]
    \begin{quantikz}
        & \qw\slice{$\rho_i$} & \gate[style={fill=red!20}]{N} & \qw\slice{$\rho_i^{\prime} = C(\rho_i)$} & \qw & \gate[wires=4, style={fill=blue!20}]{U}\slice{$\rho''_i = \rho_{i+1}$} & \qw & \qw \\
        & \qw & \gate[style={fill=red!20}]{N} & \qw & \qw & \qw & \qw & \qw \\
        & \qw & \gate[style={fill=red!20}]{N} & \qw & \qw & \qw & \qw & \qw \\
        & \qw & \qw & \qw & \gate[style={fill=blue!20}]{R} &\qw & \qw & \qw
    \end{quantikz}

\caption{The schematic picture of one iteration of HBAC. $\rho_i$ is a density matrix of $n+1$ qubits which the first one is the target qubit and the last one is the reset qubit. 
In a realistic implementation, there is also noise. 
We consider the uncorrelated noise as an extra operation in each iteration which is demonstrated by $N$. 
During the reset process, the noise channels are applied to the computational qubits. $ \rho^\prime_i = C(\rho_{i})$ represents the state after effect of the quantum noise. Then the reset and compression steps make one HBAC iteration and $\rho_{i+1}= U^{\dagger}(\text{Tr}_{\text{R}}(\rho^\prime_i) \otimes \rho_{\text{R}})U$. The reset step includes the partial trace over the reset qubit and replacement of the reset qubit with $\rho_{\text{R}}$ which is in the thermal state of the reset element in contact with the heat-bath. $U$ is the compression operation.
For TSAC, this is given by unitary operator $U_{\text{TS}}$ in Eq. (\ref{eq:U_TSAC}). For PPA, $U$ represents the sort operator that is changing from one iteration to the other.}
\label{fig:1}
\end{figure}

The setting for HBAC is as follows. We consider an ensemble of $n + 1$ qubits, the first $n$ of them are referred to as the computation qubits and the last one is referred to as the reset qubit. Our results can be generalized to multi-qubit reset. 
The system is described by a density matrix in $\mathcal{H}_{\text{C}} \otimes \mathcal{H}_{\text{R}}$ Hilbert space. The computation qubits (or a subset of them) are the target elements that are to be cooled or purified.

Each iteration of HBAC involves two steps. First, there is the reset which refreshes the reset qubit to its initial state, a.k.a the reset state.  Often the reset step is the thermal equilibration of the reset qubit in contact with the heat-bath. This transfers the accumulated entropy during the previous iteration from the reset qubit to the heat-bath. Mathematically, this process is described by the following quantum channel 
\begin{equation}
R(\rho) = \text{Tr}_R(\rho) \otimes \rho_R, 
\label{eq:reset}
\end{equation}
where $\rho$ is the density matrix of the $n+1$ qubits and  
$\text{Tr}_R$ is the partial trace over the reset element. 
$\rho_R$ is the reset state, which can be expressed as 
\begin{equation}
\rho_{R} = \frac{1}{e^{\varepsilon_0} + e^{-\varepsilon_0}}\begin{pmatrix}
    e^{\varepsilon_0} & 0 \\
    0 & e^{-\varepsilon_0}
  \end{pmatrix}, 
  \label{eq:reset_state}
\end{equation}
where $\varepsilon_0$ is referred to as the polarization of the reset qubit. This is inspired by the thermal distribution given by the Boltzmann distribution where $\varepsilon_0$ depends linearly on the energy gap of the qubit and inversely on the temperature.

For an arbitrary qubit state, the polarization can be characterized in terms of the eigenvalues of the density matrix. We define $\Lambda(\rho)$ to denote the largest eigenvalue of the qubit state $\rho$. The polarization can be expressed as 
\begin{equation}
  \varepsilon = \frac{1}{2} \log({\frac{\Lambda(\rho)}{1-\Lambda(\rho)}}).
  \label{eq:polarization}
\end{equation}
Both polarization and $\Lambda(\rho)$ can be  used as measures of purity. Polarization is commonly used in the NMR and HBAC literature. Here for some of our results, we will be using $\Lambda(\rho)$ directly.

In each iteration of HBAC, the reset step is followed by the compression which is done by the application of a unitary operation over the $n+1$ qubits. For PPA, this is the sort operation, i.e. a permutation operation that sorts the diagonal elements of the density matrix decreasingly. For TSAC, the compression operation is the following operation that was introduced in \cite{raeisi2019novel}.
\begin{equation}
   U_{\text{TS}} = \begin{pmatrix}
    1   &   &   &   &  \\
    &  \sigma_{X}  &  &   &   \\
    &   &  \ddots  &  & \\
    &   &   &   \sigma_{X} & \\
    &   &   &   &   1
  \end{pmatrix},
   \label{eq:U_TSAC}
\end{equation}
where $\sigma_{X}$ is the Pauli X matrix. Putting the reset and compression together, each iteration of HBAC can be described as  
\begin{equation}
   \rho \rightarrow \rho^\prime= U^{\dagger}(\text{Tr}_{R}(\rho) \otimes \rho_{R})U,
   \label{eq:one_iteration}
\end{equation}
with $U$ the compression operation.

Both PPA and TSAC converge to the same asymptotitc state that gives the HBAC limit. For the first qubit, the limit for the polarization is \cite{PhysRevLett.114.100404, raeisi2019novel}
\begin{equation}
   \varepsilon = 2^{n - 1} \varepsilon_{0}.
   \label{eq:HBAC_limit}
\end{equation}

For ideal HBAC, it is assumed that the reset element is  strongly coupled to the heat-bath and is  equilibrated faster than the computation qubits. This means that the reset element can be reset faster than the computation element. 

Typically, it is also assumed that the state of the computation qubits remains unchanged while the reset qubits are reset. 
This is the key assumption that we are revisiting and relaxing in this work. In particular, we take into account that during the reset process, the computation qubits could undergo some noise.

Note that there are two main assumptions in HBAC that need to be revisited. The first one is that the compression operation is noiseless. The second one is that while the reset takes place, the computation qubits remain unchanged. 
Here we focus on the second assumption and include the single-qubit noise on the computation elements during the reset. This choice is mostly motivated by the experimental feasibility of the two assumptions. For instance, for a spin implementation, the unitary operations can be implemented with high fidelity. But for the reset, the computation qubits do equilibrate, although it may not be as much as the reset qubit. Therefore this work is mainly focused on including the relaxation of the computation qubits. In particular, we consider the generalized amplitude damping (GAD) channel. We also investigate other single-qubit noise channels such as the depolarizing and bit-flip noise. 

Fig.~\ref{fig:1} shows how we consider noise. The noise channels are depicted in red. The noise occurs in parallel to the reset. In each iteration, during the reset, the computation qubits undergo some noise. Note that similar to the reset operation, the noise cannot be described by unitary quantum gates. For most of this work, we consider the GAD for the noise. We included other noise channels in the appendix \ref{HBAC_Depolarizing_Bitflip}.

The single-qubit noise channels can be represented by
\begin{equation}
    C(\rho) = \sum_{k=1}^{m} E_k \rho E_k^{\dagger},
    \label{Eq:single_qubit_noise}
\end{equation}
where $\{E_k\}_{k = 1}^{m}$ are Kraus operators for the noise \cite{nielsen_chuang_2010}. The Kraus operators for the GAD are specified with two parameters, $p$ and $\gamma$. The explicit form of the Kraus operators are given in the appendix \ref{Noise_channels_Kraus}. Fig.~\ref{fig:2} shows schematically how a noise channel affects a qubit state. For the GAD noise, $\alpha$ and $\beta$ are $\gamma (1-p)$ and $\gamma p$ respectively.

Here we only consider an uncorrelated noise. This means that the noise in Eq.~\ref{Eq:single_qubit_noise} is applied to all the $n$ qubits independently. For the $n$ qubits, this leads to the following noise channel
\begin{equation}
    C(\rho) = \sum_{j=1}^{m^n} \Tilde{E}_j\,\rho\,{\Tilde{E}_j}^{\dagger},
    \label{Eq:n_qubit_noise}
\end{equation}
where $\Tilde{E}_j = E_{k_1} \otimes E_{k_2} \otimes \dots \otimes E_{k_n}$ and $\{E_{k_i}\}_{k_i = 1}^{m}$ are single-qubit Kraus operators.

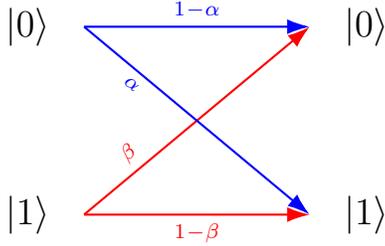
\begin{figure}
\setlength{\unitlength}{0.20mm}
\begin{tikzpicture} 
    \tikzset{>={Latex[width=2mm,length=3mm]}}
    \path (0,0) node(1_left) {\Large $\ket{1}$};
    \path (0,2.5) node (0_left) {\Large $\ket{0}$}; 
    \path (4.5,0) node (1_right) {\Large $\ket{1}$}; 
    \path (4.5,2.5) node (0_right) {\Large $\ket{0}$}; 
    \draw[->,thick,red] (0.75,0) -- node[below] {\footnotesize $1\!-\!\beta$} (3.75,0);
    \draw[->,thick,blue] (0.75,2.5) -- node[above] {\footnotesize $1\!-\!\alpha$} (3.75,2.5);
    \draw[->,thick,red] (0.75,0) -- node[near start,above, sloped] {\footnotesize $\beta$} (3.75,2.5);
    \draw[->,thick,blue] (0.75,2.5) -- node[near start,below, sloped] {\footnotesize $\alpha$} (3.75,0);
\end{tikzpicture}%
\caption{The schematic picture of the effect of a single-qubit noise channel. The noise flips the state of a qubit from $\ket{0}$ to $\ket{1}$ with probability $\alpha$ and flips the state from $\ket{1}$ to $\ket{0}$ with probability $\beta$. 
The $\alpha$ and $\beta$ for the generalized amplitude damping noise are $\gamma (1-p) $ and $ \gamma p $ respectively.} 
\label{fig:2}
\end{figure}

\begin{figure*}
\includegraphics[scale=0.315]{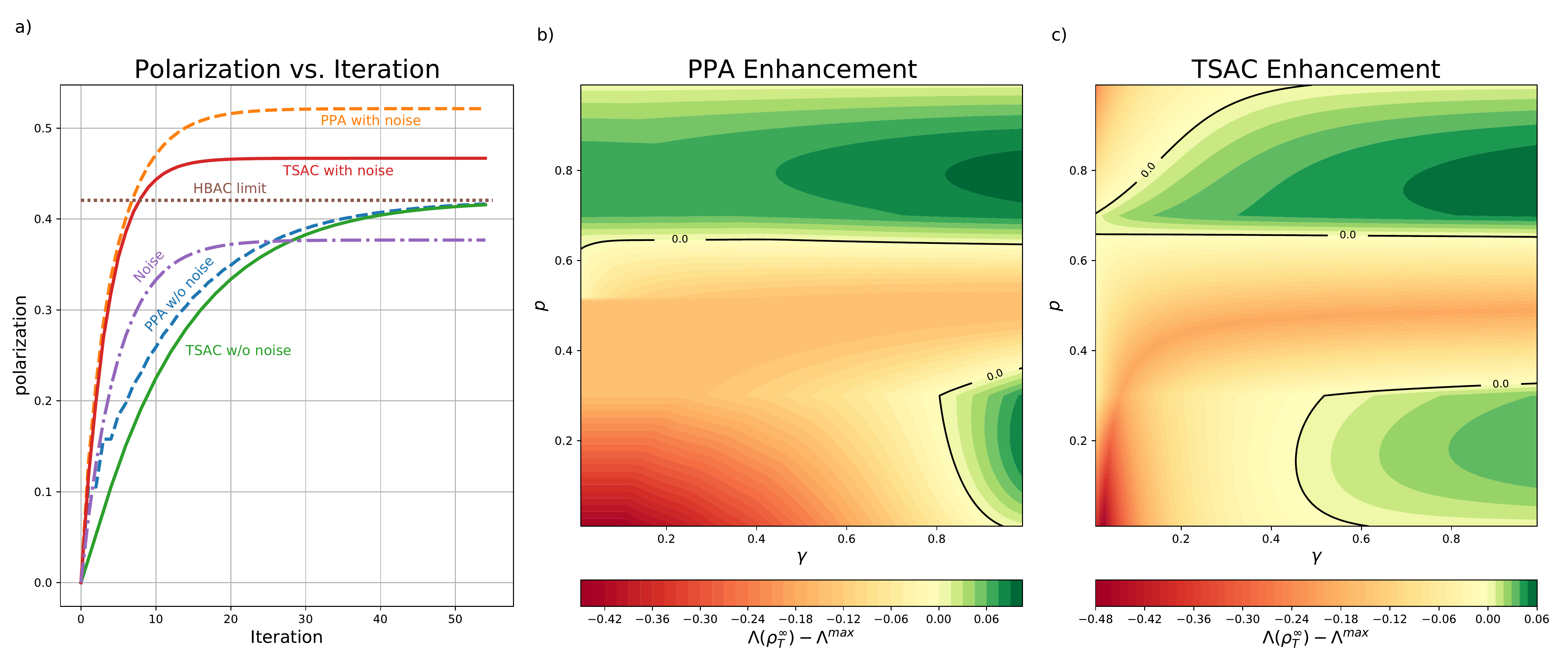}
\caption{The simulation results for noisy HBAC with generalized amplitude damping noise with $n=3$ computational qubits and $\varepsilon_0 = 0.11$. (a) shows polarization of the target qubit vs. the number of iterations. This is for $\gamma = 0.20$ and $p = 0.68$. 
(b) and (c) show the purity enhancement, $E = \Lambda(\rho_T^{\infty}) - \Lambda^{\text{max}}$ versus $p$ and $\gamma$ for PPA and TSAC respectively. Note that $\Lambda^{\text{max}}$ is the maximum of the largest eigenvalue of the asymptotic density matrix of the target qubit for both the HBAC and GAD.  In (b) and (c) the green region illustrates the positive $E$ where the asymptotic polarization of the noisy HBAC is higher than the asymptotic polarizations of HBAC and GAD individually. We refer to this region is as the green zone.}
\label{fig:3}
\end{figure*}

We analyze both PPA and TSAC and numerically simulate them under the GAD noise for the full range of the noise parameters, $p$ and $\gamma$. We do these simulations for different numbers of computation qubits $n$ and different values of the reset polarization $\varepsilon_0$. 
For the noisy PPA, the process in Fig.~\ref{fig:1} is simulated with $U$, the sort operation. The sort operation changes in each iteration and it is not feasible to carry out this analysis analytically. 
However, the noisy TSAC, similar to the noiseless process,  makes a time-independent Markov chain, i.e. a constant transition matrix takes $\rho^{t}$ to $\rho^{t+1}$. Similar to the analysis in \cite{raeisi2019novel}, we find the transition matrix of the Markov process. 
Without loss of generality we only consider diagonal density matrices in our analysis (for more details see the appendix \ref{Non_diagonal_density_matrices}). For the density matrix of the computational qubits in the $t$th iteration, the diagonal elements can be written as a vector with $2^n$ elements
\begin{equation}
   \vec{p^{t}} = \{ p^t_1 , p^t_2 , \dots , p^t_{2^{n}}\}. \label{Eq:computational_densitymatrix}
\end{equation}

The effect of both TSAC and the GAD noise on the diagonal elements is given by
\begin{equation}
   \vec{p^{t+1}} = T_{\text{TSAC}} \; T_{\text{GAD}} \; \vec{p^t} = T \;  \vec{p^t},\label{Eq:computational_changes}
\end{equation}
where $T_{\text{TSAC}}$ and $T_{\text{GAD}}$ are the transition matrices for TSAC and GAD respectively. 
The explicit form of $T_{\text{TSAC}}$ and $T_{\text{GAD}}$ are given in the appendix \ref{Noise_channels_transfer_matrix}. 
We numerically find  the eigenvalues of the full transfer matrix, i.e. $T = T_{\text{TSAC}}  T_{\text{GAD}}$. The spectrum of the full transfer matrix has a single eigenvector of eigenvalue one and all the other eigenvalues have magnitudes less than one. This indicates that for the noisy Markov process, the $\vec{p}$ converges to the eigenvector corresponding to the eigenvalue one.

For both PPA and TSAC, our results show that the noisy HBAC converges. 
We also find that for certain values of the noise parameters, the noisy HBAC outperforms the ideal HBAC and exceeds the cooling limit of HBAC. Fig.~\ref{fig:3}-a shows the simulation results for both PPA and TSAC with and without noise. 
For Fig.~\ref{fig:3}-a, the simulations are for $n=3 $ computation and one reset qubits and the reset polarization is $\varepsilon_0 = 0.11$. 
Also, we are considering the GAD noise with parameters $p=0.68$, $\gamma = 0.20$. 
Fig.~\ref{fig:3}-a  shows that for this specific set of parameters, both for PPA and TSAC, noisy HBAC converges to a higher polarization, compared to the noiseless HBAC limit. Note that the asymptotic polarization is also higher than the asymptotic polarization of the GAD noise alone. 

We identify the region in the parameter space where the noise enhances the HBAC as the ``green zone''. Note that for $\alpha=0$ and $\beta=1$, the noise would fully purify the qubit, even without the HBAC, but this is not of interest because the enhancement is due to the unrealistic and extreme noise channel. We exclude these regions from the green zone. 
To be more specific, the green zone specifies the set of values of the parameters $p$ and $\gamma$ for which, the asymptotic polarization of the first qubit surpasses both the noiseless HBAC or the noise channel individually. In other words, for the parameters in the green zone, the combination of the GAD noise and HBAC is stronger than noise or HBAC alone. 
Mathematically, the green zone is identified by the region in which

\begin{align}
    \Lambda(\rho_T^{\infty}) \geq \Lambda^{\text{max}}&, 
    \nonumber
    \\
    \Lambda^{\text{max}} =  \text{max} \{\Lambda(\rho_{T,\text{HBAC}}^{\infty}), & \Lambda(\rho_{T,\text{GAD}}^{\infty})\},
    \label{Eq:green_zone}
\end{align}
where $\Lambda(\rho_T^{\infty})$ is the largest eigenvalue of the asymptotic density matrix of the target qubit in the noisy HBAC process.

To quantify the enhancement, we use the difference between $\Lambda(\rho_T^{\infty})$ and the $\Lambda^{\text{max}}$.  
More specifically, we introduce ``purity enhancement'' as $E = \Lambda(\rho_T^{\infty}) - \Lambda^{\text{max}}$. 
A positive purity enhancement indicates that the noisy HBAC is outperforming the HBAC and the noise individually. 
The green zone identifies the sets of parameters where the enhancement is positive. 
Fig.~\ref{fig:3}-b and c show the purity enhancement for different values of $p$ and $\gamma$ for PPA and TSAC respectively. The positive enhancement is depicted in green and the color bar shows the value of the enhancement. 

\begin{figure}
\includegraphics[scale=0.35]{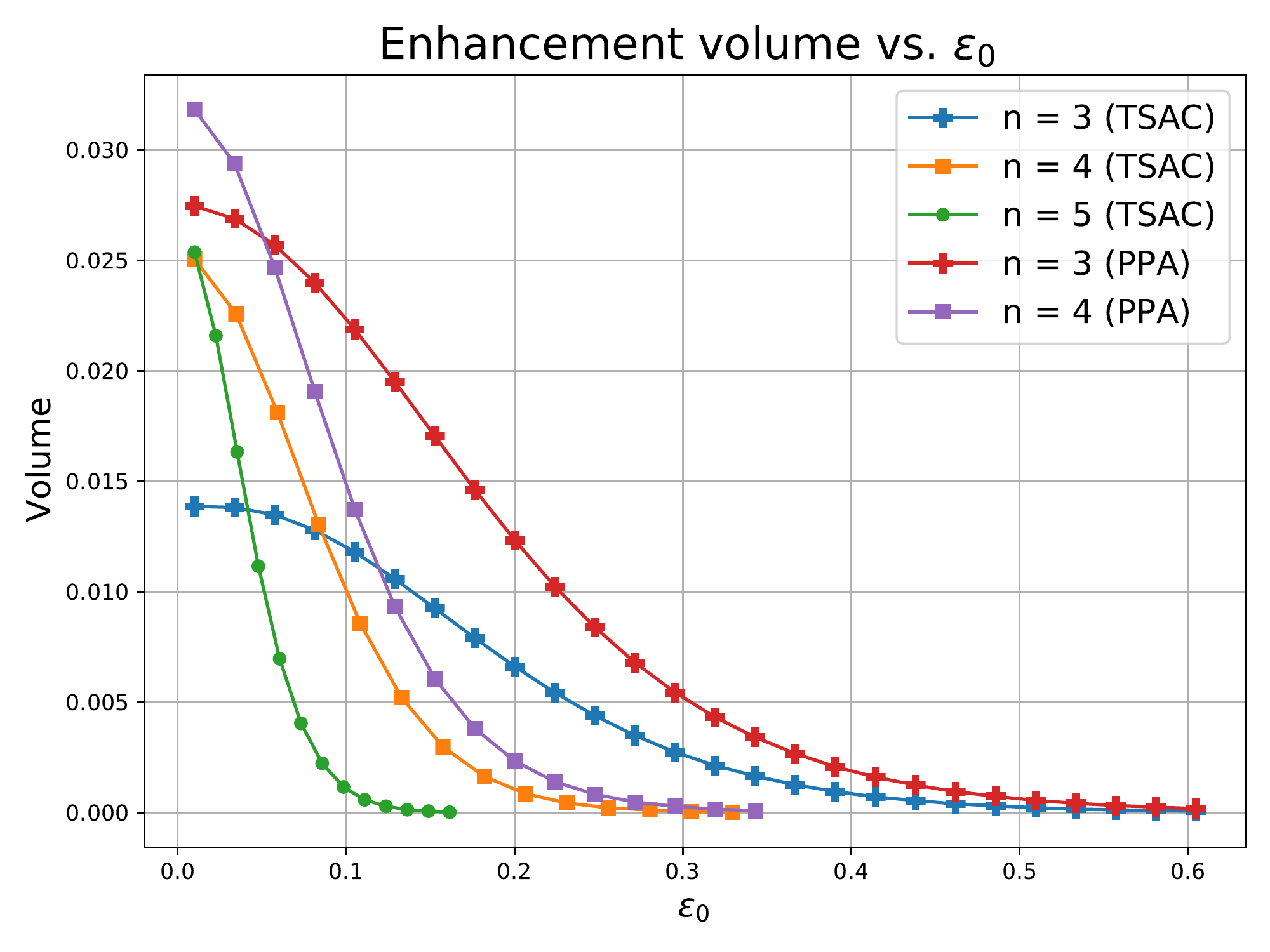} 
\caption{Enhancement volume versus the reset polarization. The volume is calculated numerically for different values of $n$ where $n + 1$ is the number of total qubits (for more details, see the appendix \ref{Enhancement_volume}).}
\label{fig:4}
\end{figure}

Both the green zone and the purity enhancement depend on the  number of qubits  and the reset polarization. For comparison, we introduce the volume of the green zone. 
This quantity depends on both the size of the green zone in the parameter space, i.e. ($p$,$\gamma$) and the amount of enhancement. This volume is calculated numerically for different $ \varepsilon_0 $ and different numbers of qubits.  
Fig.~\ref{fig:4} shows how this volume changes with the reset polarization and for different numbers of qubits. 
The details of the numerical estimation are explained in the appendix \ref{Enhancement_volume}. 

Fig.~\ref{fig:4} shows that the enhancement is more pronounced for smaller values of reset polarization. Note that for large enough values of the reset polarization where the HBAC converges to a pure state, there is no room for enhancement. Similarly, the HBAC limit grows as the number of qubits $n$ increases, which results in a sharp reduction of the green zone with increasing the reset polarization. 
It is also evident from the plot that the GAD noise enhances PPA more effectively compared to TSAC.

Next, we investigate how the noise can enhance the performance of HBAC. 
To this end, we assume that we are starting with the asymptotic state of HBAC and we consider one iteration of the noisy HBAC. 
We refer to this state as the optimal asymptotic state (OAS). 
Note that when we are starting with OAS, the HBAC alone would not change the state. Also, the noise alone would reduce the polarization. We show that for specific values of the noise parameters, the combinations of HBAC and GAD noise would increase the purity of the state beyond the HBAC limit.

The OAS has been established in \cite{PhysRevLett.114.100404} and is given by a diagonal density matrix with the following vector for the diagonal elements
\begin{equation}
    \vec{p}_{\text{OAS}} = p_0 \{1, e^{-2 \varepsilon_0}, e^{-4 \varepsilon_0}, \dots, e^{-2 (2^n - 1) \varepsilon_0}  \},
\label{Eq:rho_HBAC}
\end{equation}
where $ p_0 $ is the normalization factor. The 
density matrix of the target qubit is obtained by tracing over all qubits except the target, i.e. the first qubit. This gives
\begin{equation}
    \rho_{\text{T}} = \{ \Lambda(\rho_{\text{T}}) , 1 - \Lambda(\rho_{\text{T}}) \},
\label{Eq:rho_HBAC_target}
\end{equation}
where $ \Lambda(\rho_{\text{T}}) $ is 
\begin{equation}
    \Lambda(\rho_{\text{T}}) = p_0 \sum_{m = 1}^{2^{(n-1)}} e^{-2 (m-1) \varepsilon_0}.
\label{Eq:target_eigenvalues}
\end{equation}

We start with OAS and let the state evolve through a single iteration as in Fig.~\ref{fig:1}. This comprises the noise, the reset and the compression. In one iteration, the variation in the polarization of the target qubit includes the changes caused by the noise and compression steps. 
\begin{figure*}
\includegraphics[scale=0.480]{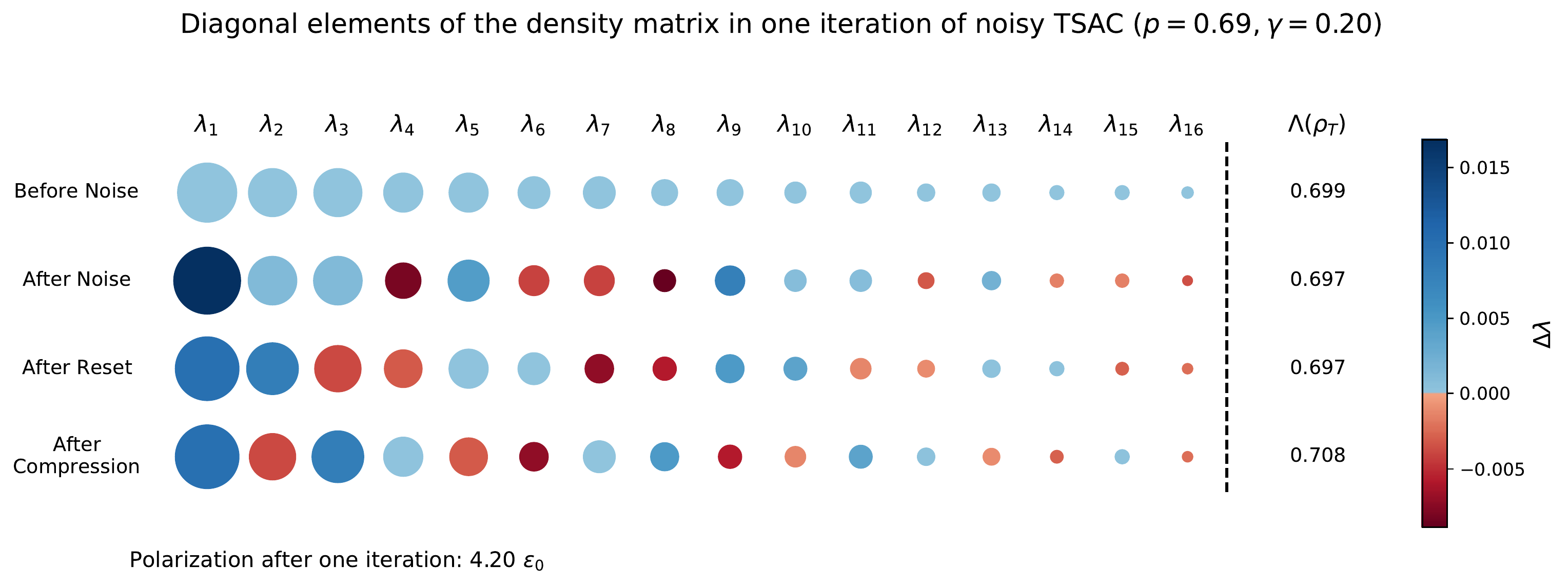} 
\caption{The results of one iteration of noisy HBAC on the asymptotic state of HBAC. This if for a 4-qubit system with $\varepsilon_0 = 0.11$,  $p = 0.69$ and $\gamma = 0.20$ for the GAD noise. The parameters are chosen such that they satisfy  Eq.~\ref{Eq:enhancement_condition}. $\{ \lambda_i \}$ 
denote the diagonal elements of the density matrix and each row represents a step in this iteration. For each element, the size and the color of circles display the value and the changes $\Delta \lambda_i$  (compared to the initial value) of that element. The initial state is the OAS state (before noise). The largest element of the density matrix of the target qubit ($\Lambda(\rho_{\text{T}})$) is displayed on the right, next to the color bar. It shows that the purity decreases after the noise, but then increases beyond the limit after the compression. This indicates that although the noise decreases the purity, it prepares the state such that it would be more affected by the compression. 
}
 
\label{fig:5}
\end{figure*}
We can analytically calculate  the state after the effect of the GAD noise. 
The density matrix of the target qubit after application of the GAD noise is
\begin{equation}
    \rho^\prime_{\text{T}} = \{ \Lambda(\rho^\prime_{\text{T}}) , 1 - \Lambda(\rho^\prime_{\text{T}}) \},
\label{Eq:rho_after_noise_target}
\end{equation}
where $\Lambda(\rho^\prime_{\text{T}}) $ is as follows 
\begin{equation}
    \Lambda(\rho^\prime_{\text{T}}) = (1 - \alpha) \; \Lambda(\rho_{\text{T}}) +  \beta \; (1 - \Lambda(\rho_{\text{T}})).
    \label{Eq:after_noise_target_eigenvalues} 
\end{equation}
After the compression, i.e. application of $ U_{\text{TS}}$, the density matrix of the target qubit changes to 
\begin{equation}
   \rho''_{\text{T}} = \{ \Lambda(\rho''_{\text{T}}) , 1 - \Lambda(\rho''_{\text{T}}) \}.
\label{Eq:rho_after_TSAC_target}
\end{equation}
In which $ \Lambda(\rho''_{\text{T}}) $ is
\begin{equation}
    \Lambda(\rho''_{\text{T}}) = \Lambda(\rho^\prime_{\text{T}}) +  \frac{e^{\varepsilon_0}}{z} \; p^\prime_{2^{n-1}+1} - \frac{e^{- \varepsilon_0}}{z} \; p^\prime_{2^{n-1}}. 
    \label{Eq:after_TSAC_target_eigenvalues} 
\end{equation}
We are looking for a noise that increases the polarization of $ \vec{p}_{\text{OAS}} $. As explained, increasing the polarization is equivalent to positive $ \Delta \Lambda(\rho_{\text{T}}) = \Lambda(\rho''_{\text{T}}) - \Lambda(\rho_{\text{T}}) $. Based on Eq.~\ref{Eq:after_noise_target_eigenvalues} and Eq.~\ref{Eq:after_TSAC_target_eigenvalues}, $ \Delta \Lambda(\rho_{\text{T}}) $ can be written as
\begin{align}
    \Delta \Lambda(\rho_{\text{T}}) &= \frac{e^{\varepsilon_0}}{z} \; p^\prime_{2^{n-1}+1} - \frac{e^{- \varepsilon_0}}{z} \; p^\prime_{2^{n-1}} \nonumber
    \\
    &+ \beta (1 - \Lambda(\rho_{\text{T}})) - \alpha \Lambda(\rho_{\text{T}}).
\label{Eq:delta_lambda}
\end{align}
The purity enhancement is identified by $\Delta \Lambda(\rho_{\text{T}}) > 0 $. This leads to
\begin{equation}
    \frac{e^{\varepsilon_0}}{z} \: p^\prime_{2^{n-1}+1} + \beta \: (1 - \Lambda(\rho_{\text{T}})) > \frac{e^{- \varepsilon_0}}{z} \: p^\prime_{2^{n-1}} + \alpha \: \Lambda(\rho_{\text{T}}).
\label{Eq:enhancement_condition}
\end{equation}

Fig.~\ref{fig:5} illustrates a situation where Eq.~\ref{Eq:enhancement_condition} holds. This is for a 4-qubit system with GAD noise with $p = 0.69$ and $\gamma = 0.20$. 
Fig.~\ref{fig:5} shows the evolution of the elements of the density matrix in each step. 
Each row represents a step in an iteration and each column shows an element of the density matrix of the system. The size of each circle indicates the value of the corresponding element, while the color shows the change in the value of that element from its initial value (OAS). 
The last column shows the largest eigenvalue of the target qubit, $\Lambda(\rho_{\text{T}})$. 
Initially, the qubits are in the OAS state (first row). The second row shows the state after the GAD noise. The right column shows that the polarization of the target qubit slightly decreases after the noise. The dark blue circles show that despite the decrease in the polarization of the target, some of the elements increase with noise. Specifically, after the noise, the first and ninth elements increase, while the seventh and eighth elements decrease.
Next, there is the reset (third row), which does not change the polarization of the target qubit. The reset  prepares the state for the compression, which is the last step.
The eighth and ninth elements remain unchanged after the reset step. The compression swaps these two which increases the sum of the first eight elements, i.e. $\Lambda(\rho_{\text{T}})$. 
This one iteration increases the polarization to $4.20 \: \varepsilon_0$, which is $0.2 \: \varepsilon_0$ more than the polarization of the optimal asymptotic state of HBAC, OAS.

In conclusion, we investigated noisy HBAC. 
Specifically, we investigated HBAC under the generalized amplitude damping channel. 
We showed that the generalized amplitude damping channel in the specific region of its parameter space can enhance the cooling process and exceed the cooling limit of HBAC.

We identified regions of the parameter space where noise enhances the HBAC. We introduced the enhancement volume to characterize the amount of enhancement as well as the area of the parameter space in which noisy HBAC exceeds the HBAC and noise individually. We showed that the enhancement  volume decreases with the reset polarization. 
We also investigated how the noise affects the process and how the noisy HBAC can outperform the HBAC. 

Note that HBAC is established under the assumption of perfect operation and noiseless qubits. This has been partly motivated by the intuition that noise would make HBAC less effective. Contrary to the common belief, we showed that noisy HBAC may be even more effective. This also seems to be consistent with experimental results in \cite{zaiser2018experimental} where in some instances, the polarization exceeded the limit of HBAC.

We studied the effect of bit-flip, phase-flip and depolarizing  channels too, however, they do not lead to any enhancement over noiseless HBAC and it is only the  generalized amplitude damping noise that can enhance the HBAC limit. 

Our results can be used to set a foundation for expanding the algorithmic cooling techniques beyond the noiseless limit. 

It is also interesting to investigate the effect of these noises from a quantum thermodynamic point of view \cite{gemmer2009quantum, vinjanampathy2016quantum}. For instance, it would be interesting to revisit some of the results on the third law of thermodynamics under imperfect and noisy conditions \cite{browne2014guaranteed, reeb2014improved, masanes2017general}. 
It is also interesting to understand the noise in the context of the resource theory of purification \cite{horodecki2003reversible, streltsov2018maximal, gour2015resource}. In particular, it is unclear if the effect of the noise in enhancing the purification is specific to algorithmic cooling or can be generalized. 

From a practical point of view, these results may be used to engineer the system in a region of the noise parameters where it is possible to exceed the cooling limit of HBAC. Or in some cases, it may be possible to exploit the noise present in the system to go beyond the limits of HBAC.

\begin{acknowledgments}
	We would like to thank Michele Mosca for fruitful discussions. This work was supported by the research 
	grant system of Sharif University of Technology (G960219). 
\end{acknowledgments}


\bibliographystyle{apsrev4-2}
\bibliography{references}



\widetext

\appendix
\section{Noise channels Kraus operators}\label{Noise_channels_Kraus}

To take account of noise in HBAC, we analyze the following noise channels. The single-qubit noise channels' Kraus operators are as follows \cite{nielsen_chuang_2010}
\begin{enumerate}
\item
Generalized amplitude damping channel:
\begin{align}
E_0&=
\sqrt{p}\begin{pmatrix}
    1 & 0 \\
    0 & \sqrt{1-\gamma}
  \end{pmatrix},
&
E_1&=
\sqrt{p}\begin{pmatrix}
    0 & \sqrt{\gamma} \\
    0 & 0
  \end{pmatrix}, \nonumber
\\
E_2&=
\sqrt{1-p}\begin{pmatrix}
    \sqrt{1-\gamma} & 0 \\
    0 & 1
  \end{pmatrix},
&
E_3&=
\sqrt{1-p}\begin{pmatrix}
    0 & 0 \\
    \sqrt{\gamma} & 0
  \end{pmatrix}.
  \label{App.Eq:1}
\end{align}
Where $ \gamma $ and $ p $ are the parameters that the generalized amplitude damping noises are identified with them.  From Fig.~\ref{fig:2} we consider that a quantum noise flips the basis of a qubit Hilbert space from $\ket{0}$ to $\ket{1}$ and vice versa, with probabilities $\alpha$ and $\beta$, respectively. This implies that the probability of remaining in $\ket{0}$ is $1-\alpha$ and for $\ket{1}$ is $1-\beta$ as well. Consequently, for the GAD noise, $\alpha = \gamma (1-p)$ and $\beta = \gamma p$ (Table.~\ref{table:1}).


\item
Depolarizing channel:
\begin{align}
    E_0&=\sqrt{1-\frac{3p}{4}} \mathds{1},
    &
    E_1&=\sqrt{\frac{p}{4}}X,\nonumber
    \\
    E_2&=\sqrt{\frac{p}{4}}Y,
    &
    E_3&=\sqrt{\frac{p}{4}}Z.
    \label{App.Eq:2}
\end{align}

For depolarizing channel, $\alpha$ and $\beta$ are the same and equal to $\frac{p}{2}$ (Table.~\ref{table:1}). 

\item
Bit-flip channel:
\begin{align}
    E_0&=\sqrt{p}\mathds{1},
    &
    E_1&=\sqrt{1 - p}X.
    \label{App.Eq:3}
\end{align}

In bit-flip channel $1 - p$ is the probability of flipping $\ket{0}$ and $\ket{1}$ to each other and they remain in their state with probability $p$ (Table.~\ref{table:1}).
\end{enumerate}

It is worth mentioning that the phase damping channel does not alter the diagonal elements of the density matrix (for more information see the appendix \ref{Non_diagonal_density_matrices}). Therefore, we do not consider the phase damping channel in our calculations. 

\section{TSAC and Noise channels' transfer matrices}\label{Noise_channels_transfer_matrix}
According to Eq.~\ref{Eq:computational_changes} the effect of both TSAC and the noise on the diagonal elements is given by $ T_{\text{TSAC}} T_{\text{Noise}}$. For a system with $n$ computational qubits, the transfer matrix of TSAC is as follows \cite{raeisi2019novel}
\begin{equation}
    T_{\text{TSAC}} = \frac{1}{z} \begin{pmatrix}
    e^{\varepsilon_0} & e^{\varepsilon_0} & 0 & \cdots & 0 \\
    e^{-\varepsilon_0} & 0 & e^{\varepsilon_0} & \cdots & 0 \\
    0 & e^{-\varepsilon_0} & 0 & \cdots & 0 \\
    0 & 0 & \cdots & \ddots & \vdots \\
    0 & 0 & \cdots & e^{-\varepsilon_0} & e^{-\varepsilon_0}
   \end{pmatrix}_{2^n \times 2^n}, 
   \label{Eq: T_TSAC}
\end{equation}
where $z = (e^{-\varepsilon_0} + e^{\varepsilon_0})$, the partition function of the reset qubit. And the noise for single-qubit is explained in the appendix \ref{Noise_channels_Kraus}. Each single-qubit state is changed with probabilities $\alpha$ and $\beta$ individually in a multiple-qubit system, due to the noise. The reason is that uncorrelated noise channels are applied in the process (Fig.~\ref{fig:1}). Accordingly, the transfer matrix of each noise channel is a $2^n \times 2^n$ matrix which its elements are given as
\begin{equation}
    T_{\text{Noise},ij} = (1-\alpha)^{d_H(\Bar{i}\wedge \Bar{j})}\: \alpha^{d_H(\Bar{i}\wedge j)}
    \: \beta^{d_H(i\wedge \Bar{j})} \: (1-\beta)^{d_H(i\wedge j)}
    \label{App.Eq:4}
\end{equation}
where $i$ and $j$ are referred to arbitrary states of a multiple-qubit system and $d_H(x)$ is the hamming distance between $x$ and $0$. For example, for $n=3$, the state $\ket{i} = \ket{010}$ transfers to the state $\ket{j} = \ket{100}$ with the probability equal to $T_{\text{Noise},ij} = (1-\alpha) \alpha \beta$. 

For the GAD, depolarizing and bit-flip channels the transfer matrix has only one eigenvalue equal to $1$, and the others' absolute values are less than $1$, as we verify this numerically. Consequently, the system converges to the eigenstate corresponding to $1$ eigenvalue.

\begin{table}
\centering
\begin{tabular}{|c|c|c|c|}
\hline
\textbf{probability} & \textbf{GAD}            & \textbf{bit-flip} & \textbf{Depolarizing} \\ \hline
$\alpha$             & $\gamma(1\!-\!p)$ & $1\!-\!p$         & $\frac{p}{2}$         \\ \hline
$\beta$              & $\gamma p$        & $1\!-\!p$         & $\frac{p}{2}$         \\ \hline
\end{tabular}
\caption{The $\alpha$ and $\beta$ for quantum noise (Fig.~\ref{fig:2}). The noise flips the basis of a single-qubit Hilbert space from $\ket{0}$ to $\ket{1}$ with probability $\alpha$ and flips the state from $\ket{1}$ to $\ket{0}$ with probability $\beta$.}
\label{table:1}
\end{table}

\section{Purity enhancement volume}\label{Enhancement_volume}

To define a measure for enhancement, we calculate the volume of the green region and integrate it numerically for fixed $\varepsilon_0$ and $n$. Specifically, we discretize the parameter space ($p$,$\gamma$) and divide it into $N_p \times N_\gamma$ segments of length $ \Delta p \times \Delta \gamma$ respectively. We set $\Delta p =\Delta \gamma = 0.01$ and $N_p = \frac{1}{\Delta p}, \;N_{\gamma} = \frac{1}{\Delta \gamma}$. Thus, the volume is calculated as follows 
\begin{equation}
    V = \sum_{i,j}^{N_p,N_{\gamma}} \Delta p \, \Delta  \gamma \max(0 , E_{i,j}),  
    \label{App.Eq:5}
\end{equation}
where $E_{i,j}$ is the purity enhancement for $ p = p_i$, the $i$th values for $p$ and $\gamma = \gamma_j$, the $j$th values for $\gamma$. Eq.~\ref{App.Eq:5} only takes into account the region where the purity enhancement is positive, i.e. the green zone and for the rest of the space, assigns zero which does not contribute to the sum.

\section{HBAC with Depolarizing and bit-flip noise channels}
\label{HBAC_Depolarizing_Bitflip}

\begin{figure}
\includegraphics[scale=0.48]{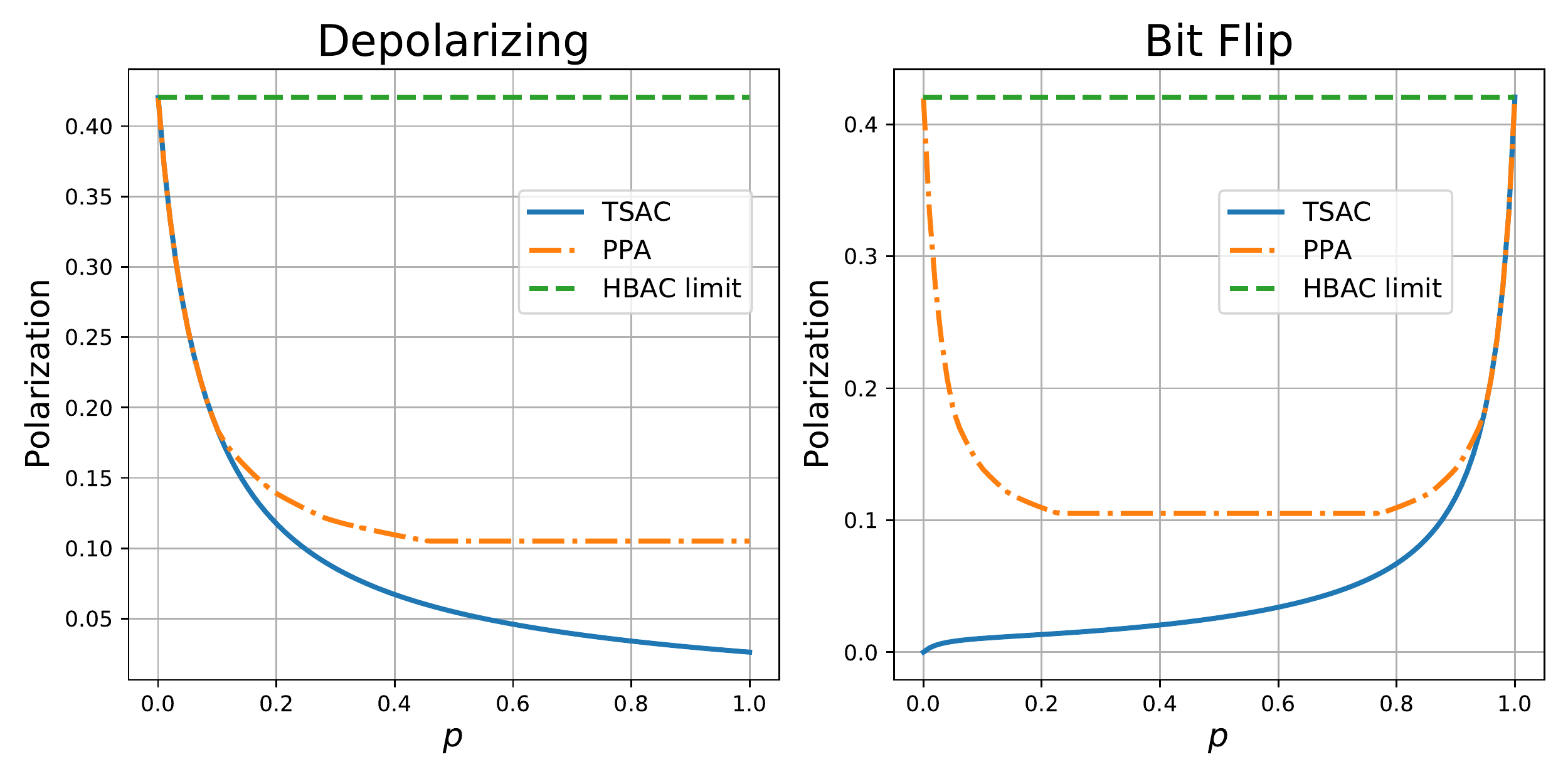}
\caption{The polarization of the target qubit versus the noise channel parameter ($p$) for TSAC and PPA with noise. The system consists of 3 computational qubits and $\varepsilon_0 = 0.11$. As shown in the figure, these channels lead to a lower polarization than
HBAC limit for every values of $p$.}
\label{fig:6}
\end{figure}

To compare the effect of other noise channels that we mentioned (appendix \ref{Noise_channels_Kraus}), we study them as well as the GAD channel. These channels are applied in the cooling process for both TSAC and PPA. In the same procedure as the GAD noise, for TSAC, we find the eigenstate corresponding to $1$ eigenvalue for each $T_{\text{TSAC}}T_{\text{Dep}}$ and $T_{\text{TSAC}}T_{\text{BF}}$. For PPA, we simulate the process until the system reaches its steady state. The results are shown in Fig.\ref{fig:6} for $n=3$ and $\varepsilon_0=0.11$. It shows that no enhancement occurs in these cases for different sets of parameters. Therefore, we are not interested in these channels in our work.

\section{Non-diagonal density matrices}\label{Non_diagonal_density_matrices}
The diagonal elements of the density matrix determine polarization, therefore only considering the changes of the diagonal elements is sufficient to calculate polarization. It was shown in \cite{raeisi2019novel} that the TSAC technique can take non-diagonal density matrices to the asymptomatic state of HBAC. In the presence of the mentioned noise channels, we show that the off-diagonal elements of the density matrix do not affect the diagonal elements and hence polarization of the first qubit.
\begin{equation}
    \rho^\prime = C(\rho) = \sum_{i = 1}^{M = m^n} E_i \rho E_i^\dagger
    \label{App.Eq:6}
\end{equation}
The diagonal element of the density matrix after the noise is \begin{equation}
    \rho^\prime_{j j} =  \sum_{i = 1}^{M} \bra{j} E_i \rho E_i^\dagger \ket{j} = \sum_{k,l = 1}^{d}
    \rho_{kl} \quad \big(  \sum_{i = 1}^{M} \bra{j} E_i \ket{l} \bra{k} E_i^\dagger \ket{j} \big ),
    \label{App.Eq:7}
\end{equation}
where $ d = 2^{n+1} $.
Since the GAD noise Kraus operators are incoherent operators (IO) (\cite{PhysRevLett.117.030401}, \cite{RevModPhys.89.041003}) we have the following equation
\begin{equation}
    (E_i^\dagger)_{k,j} (E_i)_{j,l} = \delta_{k,l}.
    \label{App.Eq:8}
\end{equation}
And the Eq.~\ref{App.Eq:7} can be simplified as 
\begin{equation}
    \rho^\prime_{j j} = \sum_{l = 1}^{d}
    \rho_{ll} \quad \big(  \sum_{i = 1}^{M} |(E_i)_{l,j}|^2 \big ).
    \label{App.Eq:9}
\end{equation} 
It shows that the diagonal elements of the density matrix after the noise $ \{ \rho^\prime_{ii} \}_{i = 1}^{d}$ only depend on the diagonal elements of the initial density matrix $ \{ \rho_{ii} \}_{i = 1}^{d}$.

\end{document}